\documentstyle[twoside]{article}

\newcount\cislo
\newcount\prvastrana
\newcount\poslednastrana
\def\publ{\hss \copyright ~Institute of Physics, SAS, Bratislava, Slovakia\hss}

\def\title#1{\begin{center}{\bf #1}\end{center}\smallskip}
\def\author#1#2{\begin{center}{\bf #1}\\ {\sl #2}\end{center}}

\def\abstract#1{\begin{quotation}\noindent{\small#1}\end{quotation}\medskip}

\def\caption#1{\noindent\small#1}

\def\ps@headings{\let\@mkboth\markboth

\def\@oddfoot{\ifnum\thepage=\prvastrana{
\noindent{\small \publ}\rm\thepage}
\else{}\fi}

\def\@evenfoot{}

\def\@evenhead{\ifnum\thepage=10000{\thepage\quad\sl{\runauthor: \nazov\hfill}}
\else{\thepage\hfil{\sl\runauthor}\qquad}\fi}

\def\@oddhead{\ifnum\thepage=\prvastrana{}
\else \qquad{\sl\shorttitle}\hfill\thepage\fi}

\def\sectionmark##1{\markboth {\uppercase{\ifnum \c@secnumdepth
>\z@
 \thesection\hskip 1em\relax \fi ##1}}{}}\def\subsectionmark##1{\markright
{\ifnum \c@secnumdepth >\@ne
 \thesubsection\hskip 1em\relax \fi ##1}}}

\def\refer#1#2#3#4#5{#1:\ {\sl #2}\ {\bf #3}\ {(#4)}\ #5;\ }

\begin{document}
\title{A Constituent Picture of Hadrons from Light-Front QCD}
\author{Robert J. Perry}{Department of Physics,
The Ohio State University, Columbus, Ohio, USA}
%
\abstract{It may be possible to derive a constituent approximation for bound
states in QCD using hamiltonian light-front field theory.  Cutoffs that violate
explicit gauge invariance and Lorentz covariance must be employed. A
similarity renormalization group and coupling coherence are used to compute
the effective hamiltonian as an expansion in powers of the canonical QCD
running coupling constant.  At second order the QCD hamiltonian contains a
confining interaction, which is being studied using bound state perturbation
theory.  Explicit constituent masses appear because of symmetry violations, and
confinement also produces mass gaps, leading to the possibility of an accurate
non-perturbative constituent approximation emerging in light-front QCD.}
%
\def\runauthor{Robert J. Perry}
\def\shorttitle{Constituent Picture in Light-Front QCD}

\vskip.15in
\noindent {\bf 1. Introduction}
\vskip.1in

The solution of Quantum Chromodynamics in the non-perturbative domain remains
one of the most important and interesting unsolved problems in physics.  QCD is
believed to be the fundamental theory of the strong interaction, but even its
definition in the non-perturbative domain is problematic.  There are many
sources of difficulty, but they can all be traced to the fact that QCD is
formulated as a theory of an infinite number of degrees of freedom that span an
infinite number of energy scales.

The basic assumption upon which all of our work is based is that it is possible
to {\it derive a constituent picture for hadrons from QCD} [1-5].  If this is
possible, non-perturbative bound state problems in QCD are approximated as
coupled, few-body Schr{\"o}dinger equations.  For a meson, we then have,

\begin{equation}
P^- \mid \Psi\rangle = {P^2_\perp +M^2 \over P^+} \mid \Psi \rangle,
\end{equation}

\noindent where,

\begin{equation}
\mid \Psi \rangle = \phi_{q\bar{q}} \mid q\bar{q} \rangle +
\phi_{q\bar{q}g} \mid q\bar{q}g \rangle + \cdot\cdot\cdot.
\end{equation}

\noindent $P^-$ is the light-front hamiltonian, $P_\perp$ is the total
transverse momentum, $P^+$ is the total longitudinal momentum, and $M$ is the
invariant mass of the state. We assume that to `leading order' a low-lying meson
can be approximated as a quark/antiquark pair, with additional quarks and gluons
producing `perturbative' corrections that can be systematically computed.

Many severe problems must be overcome to arrive at this formulation
of the bound state problem; however, the final advantages
are huge.  The result is a formulation of the non-perturbative problem in a
form directly accessible to physical intuition, which has proven essential for
guiding approximations in atomic calculations.  Variational methods and large
matrix diagonalization are powerful numerical tools that can be used after the
hamiltonian is determined.

I must emphasize that {\it it is not our intent to simply force the constituent
approximation on the theory by employing a Tamm-Dancoff truncation on the number
of particles ab initio.}  We worked on such an approach initially [6], and
gained valuable insights; but it became clear that we have no good method of
controlling the nonlocalities resulting from particle number truncation without
a dynamical mechanism that naturally limits the number of particles in a state.

Any student of field theory should immediately be suspicious of the possibility
that a constituent approximation can arise, although QED provides an important
accepted example that guides much of our work [2].  How can a constituent
approximation arise in any field theory?

Fock space is extremely large, an infinite sum of cross products of infinite
dimensional Hilbert spaces.  It is not obvious that the low-lying eigenstates
should have significant support only in the few-body sectors of Fock space.  In
fact, this simply does not happen in perturbation theory.  In perturbation
theory high-energy many-body states do not decouple from low-energy few-body
states.  Consider an electron mixing with high-energy electron/photon states. 
The error made by simply throwing away the high energy components of the state
is infinite.  Moreover, there are an infinite number of scales and both the
electron and photon that `dress' the low-energy bare electron are in turn
dressed by additional pairs, {\it ad infinitum}.

The lesson here is quite old.  Without {\it regularization and renormalization}
a constituent picture is impossible.  Renormalization may allow us to move the
dynamical effects of high-energy, many-body states from the eigenstate to
effective interactions between effective quarks and gluons.

Low-energy many-body states also do not decouple from low-energy few-body
states.  In fact, it is common lore that hadrons are excitations on an extremely
complicated vacuum. Students of QCD expect the infinite-body vacuum to be an
integral part of every hadron eigenstate.  This is the problem that leads us to
use light-front coordinates, just as it motivated the use of the infinite
momentum frame for the formulation of the parton model.  In light-front
coordinates physical particle trajectories satisfy the kinematic relativistic
constraint

\begin{equation}
p^+ \ge 0 \;,
\end{equation}

\noindent because all velocities are equal to or less than the
velocity of light.  Since longitudinal momentum is conserved, the only
states that can mix with the zero momentum bare vacuum are those in
which every bare parton has identically zero longitudinal momentum.
For a free particle of mass $m$, the light-front energy is

\begin{equation}
p^-={ {\bf p}_\perp^2+m^2 \over p^+} \;.
\end{equation}

\noindent This energy diverges as $p^+$ approaches zero, which must happen as
the number of particles grows for fixed total longitudinal momentum.  Thus, in
light-front coordinates all many-body states become high energy states, leading
us back to the original problem of replacing the effects of high energy states
with effective interactions. This argument is naive, but there is little profit
in elaborating further at this point.

Finally, manifest gauge invariance and manifest covariance apparently require
all states to contain an infinite number of particles.  This is most easily seen,
for example, by considering rotation operators.  Rotations are dynamical in
light-front coordinates and the generators contain interactions that change
particle number.  No state with a finite number of particles transforms
correctly under rotations.  We use {\it cutoffs that violate these symmetries},
which must then be repaired by effective interactions that remove all cutoff
dependence.  The constituent approximation is possible only if these symmetries
are also treated approximately.  Proposing the violation of manifest gauge
invariance is heresy in the QCD community, but heresy sometimes leads
to progress in science.

There is a long list of questions concerning how a constituent approximation can
arise in QCD, but I mention only one.  How can confinement emerge without a
complicated vacuum?  Since we have a hamiltonian, we can use a variational
calculation to study what happens as a quark/antiquark pair are separated to
infinity.  Since the addition of gluons can only lower the energy, we must find
that

\begin{equation}
\langle \phi_{q\overline{q}} \mid H \mid \phi_{q\overline{q}} \rangle
\longrightarrow  \infty  \;\; {\rm as} \;\;R \rightarrow \infty \;.
\end{equation}

\noindent Here {\it R} is the quark separation, and the only way this matrix
element can diverge is if the hamiltonian contains a two-body interaction that
diverges.  We will see below that this apparently happens.

This discussion is not intended to convince the skeptical reader that a
constituent approximation is valid.  However, {\it the assumption that a
constituent picture emerges from QCD provides strong guidance.} A hamiltonian
approach is indicated.  Cutoffs that limit mixing of high and low energy states
are required, and they must violate explicit rotational covariance and gauge
invariance.  All non-perturbative effects attributed to the vacuum in other
approaches must directly appear in few-body effective interactions.

\vskip.15in
\noindent {\bf 2. Light-Front Renormalization Group}
\vskip.1in

The renormalization of the hamiltonian and all other dynamical observables
begins with the observation that no physical result can depend on the cutoff. In
the Schr{\"o}dinger equation,

\begin{equation}
P_\Lambda^- \mid \Psi_\Lambda \rangle = {P^2_\perp+M^2 \over P^+} \mid
\Psi_\Lambda \rangle \;,
\end{equation}

\noindent the eigenvalue, $M$, cannot depend on the cutoff.  The hamiltonian,
$P^-$, must depend on the cutoff, as does the eigenstate.  Wilson's
renormalization group was formulated starting with the observation that physical
matrix elements cannot depend on the cutoff, and we have adapted his approach to
the light-front problems we face [7].

It is not possible to discuss cutoff-independence if the cutoff is fixed, so the
central operator in Wilson's renormalization group is a transformation that
lowers the cutoff.  Given a transformation, $T$, that lowers the cutoff by a
factor of $1/2$, for example, we can define a renormalized hamiltonian to be one
which has a finite cutoff but results from an infinite number of transformations. 
The transformation determines what operators must be precisely controlled for
this limit to exist.  Near a fixed point ({\it i.e.}, a hamiltonian that does not
change under the transformation), these operators can be classified as relevant
and marginal.

In the perturbative regime relevant and marginal operators are determined by
their naive engineering dimension.  In light-front field theory there is no
longitudinal locality, only transverse locality, so it is the transverse
dimension of an operator that determines its classification.  However, while
there are a finite number of relevant and marginal operators in equal-time field
theory, the violation of longitudinal locality in light-front field theory
implies that ratios of longitudinal momenta can appear, allowing entire
functions of longitudinal momentum fractions to appear in each relevant and
marginal operator.  At first sight this appears to be a disaster; however, one
paradox we faced above was how complicated interactions associated with
non-perturbative effects such as confinement could arise in few-body
operators.  This is possible because of the violation of longitudinal locality.

To develop a light-front renormalization group we must decide what cutoff to
implement and then derive a transformation that runs this cutoff.  It is
possible to use a cutoff on the total invariant-mass of states, as is commonly
done in DLCQ calculations for example; however, such cutoffs lead to strong
spectator dependence and small energy denominators appear in the resultant
effective interactions.  We use a {cutoff on the change in free energy.}  If the
hamiltonian is viewed as a matrix such a cutoff limits how far off diagonal
matrix elements can appear. 

There is not enough space to elaborate the transformation that runs this cutoff,
so I refer the reader to the literature [1,2,4,8,9].  The transformation is
unitary, leading to what G{\l}azek and Wilson call a {\it similarity
renormalization group} [8,9].   Let $H=h_0+v$, where $h_0$ is a free
hamiltonian with $h_0 \mid \phi_i \rangle = E_{0i} \mid \phi_i \rangle$, and $v$
is cut off so that

\begin{equation}
\langle \phi_i | v | \phi_j \rangle = 0 \;,
\end{equation}

\noindent if $|E_{0i}-E_{0j}| > \Lambda$.  If this cutoff is lowered to
$\Lambda'$, the new hamiltonian matrix elements to ${\cal O}(v^2)$ are

\begin{eqnarray}
H'_{ab} &=&
\langle \phi_a|h_0+v|\phi_b\rangle 
- \sum_k v_{ak} v_{kb} \Biggl[
{\theta\bigl(|\Delta_{ak}|-\Lambda' \bigr)
\theta\bigl(|\Delta_{ak}|-|\Delta_{bk}|\bigr) \over E_{0k}-E_{0a} }
\nonumber \\ && \;\;\;\;\;\;\;\;\;\;\;\;\;\;\;\;\;\;\;\;\;\;\;\;\;\;\;
\;\;\;\;\;\;\;\;\;\;\;\;\;\;\;\;\;\;\;\;
+ {\theta\bigl(|\Delta_{bk}|-\Lambda'\bigr)
\theta\bigl(|\Delta_{bk}|-|\Delta_{ak}|\bigr) \over E_{0k}-E_{0b} }
\Biggr] ,
\end{eqnarray}

\noindent where $\Delta_{ij}=E_{0i}-E_{0j}$ and $|E_{0a}-E_{0b}|<
\Lambda'$.  To follow the details of the discussion it is important to
remember that there are implicit cutoffs in this expression because
the matrix elements of $v$ have already been cut off so that
$v_{ij}=0$ if $|E_{0i}-E_{0j}|>\Lambda$.

It is rather easy to understand this result qualitatively.  We have removed the
coupling between degrees of freedom whose free energy difference is between
$\Lambda'$ and $\Lambda$, so the effects of these couplings are forced to appear
in the new hamiltonian as direct interactions.  To first order, the new
hamiltonian is the same as the old hamiltonian, except that couplings of states
with energy differences between $\Lambda'$ and $\Lambda$ are now zero.  To second
order, the new hamiltonian contains a new interaction which sums over the
second-order effects of couplings that have been removed.  The second-order term
in the new hamiltonian resembles the expression found in second-order
perturbation theory, which is not surprising since the new hamiltonian must
produce the same perturbative expansion for eigenvalues, cross sections, etc. as
the original hamiltonian.

Equation (9) shows how the hamiltonian changes when the cutoff is lowered, and
the next step is to determine from this change what hamiltonians can emerge from
an infinite number of transformations.  The simplest result is a {\it fixed
point} hamiltonian, one which does not change under the transformation.  In
$3+1$ dimensions the only known fixed points are free field theories.  {\it
Coupling coherence} is a generalization of the fixed point idea [2,7,10].  A
coupling coherent hamiltonian reproduces itself in form, but one or more
couplings run while all additional couplings are invariant functions of these
running couplings.  For example, in QCD the canonical coupling runs at third
order.  To second order in this coupling, all interactions must reproduce
themselves exactly, with $\Lambda \rightarrow \Lambda'$.  It is not trivial to
implement this simple-sounding constraint, but at each order it determines the
hamiltonian. In all calculations to date the resultant hamiltonian is unique,
and all broken symmetries are restored to the order at which the hamiltonian is
fixed.

To second order a generic coupling coherent hamiltonian that contains $v$ must
also contain,

\begin{eqnarray}
H_{ab} &=&
\langle \phi_a|h_0+v|\phi_b\rangle 
- \sum_k v_{ak} v_{kb} \Biggl[
{\theta\bigl(|\Delta_{ak}|-\Lambda \bigr)
\theta\bigl(|\Delta_{ak}|-|\Delta_{bk}|\bigr) \over E_{0k}-E_{0a} }
\nonumber \\
&&\;\;\;\;\;\;\;\;\;\;\;\;\;\;\;\;\;\;\;\;\;\;\;\;\;\;\;\;\;\;\; 
\;\;\;\;\;\;\;\;\;\;\;\;\;\;\; +
{\theta\bigl(|\Delta_{bk}|-\Lambda\bigr)
\theta\bigl(|\Delta_{bk}|-|\Delta_{ak}|\bigr) \over E_{0k}-E_{0b} }
\Biggr] ,
\end{eqnarray}

\noindent or

\begin{eqnarray}
H_{ab} &=&
\langle \phi_a|h_0+v|\phi_b\rangle 
+ \sum_k v_{ak} v_{kb} \Biggl[
{\theta\bigl(\Lambda-|\Delta_{ak}| \bigr)
\theta\bigl(|\Delta_{ak}|-|\Delta_{bk}|\bigr) \over E_{0k}-E_{0a} }
\nonumber \\
&&\;\;\;\;\;\;\;\;\;\;\;\;\;\;\;\;\;\;\;\;\;\;\;\;\;\;\;\;\;\;\; 
\;\;\;\;\;\;\;\;\;\;\;\;\;\;\;
+ {\theta\bigl(\Lambda-|\Delta_{bk}| \bigr)
\theta\bigl(|\Delta_{bk}|-|\Delta_{ak}|\bigr) \over E_{0k}-E_{0b} }
\Biggr] .
\end{eqnarray}

\noindent Note that $v$ in these expressions is the same as that
above only to first order.  The coupling coherent interaction in $H$
is written as a power series in $v$ which reproduces itself under the
transformation, except the cutoff changes.  In higher orders the
canonical variables also run.

The light-front similarity renormalization group and coupling coherence fix the
QCD hamiltonian as an expansion in powers of the running canonical coupling.

\vskip.15in
\noindent {\bf 3. QCD: A Strategy for Bound State Calculations and Confinement}
\vskip.1in

While realistic calculations will no doubt require a more elaborate procedure,
a relatively simple strategy for doing bound state calculations can now be
outlined [2,4].

\begin{center}

\parbox{4.5in}{i)Start with the canonical hamiltonian, $H_{can}$, and use the
similarity renormalization group and coupling coherence to compute,}

\begin{equation}
H^\Lambda = h_0^\Lambda+g_\Lambda h_1^\Lambda + g_\Lambda^2 h_2^\Lambda + \cdot
\cdot \cdot
\end{equation}

\parbox{4.5in} {Truncate this series at a fixed order.}

\vskip.1in

\parbox{4.5in} {ii) Choose an approximate hamiltonian that can be treated
non-perturbatively,}

\begin{equation}
H^\Lambda = H_0^\Lambda + V^\Lambda \;.
\end{equation}

\parbox{4.5in} {You must choose $\Lambda$ and $H_0^\Lambda$ to minimize errors.}

\vskip.1in

\parbox{4.5in} {iii) Accurately solve $H_0^\Lambda$ as the leading approximation.}

\vskip.1in

\parbox{4.5in}{iv) Compute higher order corrections from $V^\Lambda$ using bound
state perturbation theory.}

\vskip.1in

\parbox{4.5in} {v) To improve the calculation further return to step (i) and
compute the hamiltonian to higher order.}

\end{center}

There are two principal reasons that this strategy will fail for QCD.  First,
the hamiltonian is computed perturbatively so that errors in the strengths of
all operators are at least as large as a power of $\alpha$.  Small errors in the
strengths of irrelevant operators tend to produce even smaller errors in
results. However, errors in marginal operators tend to produce errors of the
same order in results and small errors in relevant operators tend to produce
exponentially large errors in results.  At the minimum we expect that we will
have to fine tune relevant operators, which means tuning a finite number of
functions of longitudinal momenta.  Second, chiral symmetry breaking operators
(where light-front chiral symmetry should be distinguished from equal-time
chiral symmetry [1]) will not arise at any order in an expansion in powers of
the strong coupling constant.  We must work in the broken symmetry phase of QCD
{\it ab initio} and insert chiral symmetry breaking operators.  Simple
arguments lead us to expect that only relevant operators need to be considered
if transverse locality is maintained, but there are no strong arguments for
transverse locality in these operators.

Despite these limitations, this strategy may be applied to the study of bound
states containing at least one heavy quark [5], as discussed by Martina
Brisudov{\'a} in these proceedings; although even here masses should be tuned,
as expected.  The strategy is conceptually simple and there are no {\it ad hoc}
assumptions.

The first step is to compute the effective QCD hamiltonian to order $\alpha$.  I
refer the reader to the literature for details on the canonical hamiltonian
[11].  The first applications of the approach are to mesons [5], and we assume
that for sufficiently small cutoffs we can choose $H_0^\Lambda$ to contain only
interactions in $H^\Lambda$ that do not involve particle production or
annihilation, as dictated by our initial assumption that a constituent picture
will arise.  This means we can first focus on operators that act in the
quark/antiquark sector.  I emphasize that all operators must be computed,
and without the confining interactions in sectors containing gluons the
entire approach would make no sense.

First consider the second-order correction to the quark self-energy. This
results from the quark mixing with quark-gluon states whose energy is above the
cutoff.  If we assume that the light-front energy transfer through the
quark-gluon vertex must be less than $\Lambda^2/{\cal P}^+$, the coupling
coherent self-energy for quarks with zero current mass is

\begin{eqnarray}
\Sigma_{\Lambda}(p)&=& {g_\Lambda^2 C_F \Lambda^2 \over 4 \pi^2 {\cal 
P}^+} \Biggl\{ \ln\Biggl({ p^+ \over \epsilon {\cal P}^+} \Biggr) - {3 
\over 4} \Biggr\} + {\cal O}(\epsilon) \;.
\end{eqnarray}

\noindent Here the quark has longitudinal momentum $p^+$, while the longitudinal
momentum  scale in the cutoff is ${\cal P}^+$.  The first and most interesting
feature of this result is that I have been forced to introduce a second cutoff,

\begin{equation}
p_i^+ > \epsilon {\cal P}^+ \;,
\end{equation}

\noindent which restricts how small the longitudinal momenta of any
particle can become.  Without this second cutoff on the loop momenta,
the self-energy is infinite, even with the vertex cutoff.
This second cutoff should be thought of as a longitudinal resolution.
As we let $\epsilon \rightarrow 0$ we resolve more and more wee
partons, and in the process we should confront effects normally
ascribed to the vacuum.  In this case the wee gluons are responsible
for giving the quark a mass that is literally infinite.  Theorists who
insist on deriving intuition from manifestly gauge invariant
calculations may find this interpretation repugnant, but within the
framework of a light-front hamiltonian calculation it is quite
natural.

This second, infrared cutoff poses a problem.  If we introduce a second cutoff,
shouldn't we introduce a second renormalization group transformation to run this
cutoff and find the new counterterms required by it?  No.  {\it I will insist
that all divergences associated with $\epsilon \rightarrow 0$ cancel exactly in
all physical results for color singlet states.}  The important question is how
can these divergences cancel so that mesons have a finite mass, and the answer
to this question leads to confinement.

A nearly identical calculation leads to the second-order self-energy of
gluons, and the dominant term goes like

\begin{equation}
{g_\Lambda^2  \Lambda^2 \over {\cal P}^+}
\ln\Biggl({ p^+ \over \epsilon {\cal P}^+} \Biggr)  \;.
\end{equation}

The quark and gluon masses are infinite, which is half of the confinement
mechanism.  In addition to one-body operators we find quark-quark, quark-gluon,
and gluon-gluon interactions.  As we lower the cutoff, we remove gluon exchange
interactions, and these are replaced by direct interactions.  The analysis of
all of these interactions is nearly identical, and I consider only the
quark-antiquark interaction.  This interaction includes two pieces,
instantaneous gluon exchange which is in the canonical hamiltonian, and an
effective interaction resulting from high energy gluon exchange.  To study
confinement we need to examine the longest range part of the total interaction,
which is a piece that diverges as longitudinal momentum exchange goes to zero. 
I outline the calculation [2,4].

High energy gluon exchange cancels part of the instantaneous gluon exchange
interaction, leaving

\begin{eqnarray}
V_{singular} &=&
 - 4 g_{\Lambda}^2 C_F \sqrt{p_1^+ p_2^+ k_1^+ k_2^+}\;
\Biggl({1 \over q^+}\Biggr)^2\;
\theta\bigl(\Lambda^2 / {\cal P}^+ - |q^-| \bigr) \;
\theta\bigl(|q^+|-\epsilon {\cal P}^+\bigr) .
\end{eqnarray}

\noindent Here the initial and final quark (antiquark) momenta are
$p_1$ and $p_2$ ($k_1$ and $k_2$), and the exchanged gluon momentum is
$q$.  The energies are all determined by the momenta, $p_1^-=p^2_{\perp
1}/p_1^+$, etc.  This part of the interaction is independent of the spins. If
$\Lambda
\approx \Lambda_{QCD}$, we expect further gluon exchange to be suppressed, and
we are left with this singular interaction between the quark and antiquark.

The next step in the analysis is to take the expectation value of this
interaction between arbitrary quark-antiquark states, $\langle \Psi_2
\mid V \mid \Psi_1 \rangle$.  If we define

\begin{equation}
Q={p_1+p_2 \over 2} \;\;,\;\; q=p_1-p_2 \;,
\end{equation}

\noindent and expand the wave functions about $q=0$, we find a divergence in the
expectation value,

\begin{eqnarray}
\langle \Psi_2|V_{singular}|\Psi_1\rangle &=&
-{g_\Lambda^2 C_F \Lambda^2 \over 2 \pi^2 {\cal P}^+} 
\log\Bigl( {1 \over
\epsilon} \Bigr) \int {dQ^+ d^2Q_{\perp} \over 16\pi^3}
\phi_2^*(Q) \phi_1(Q).
\end{eqnarray}

Unless $\phi_1$ and $\phi_2$ are the same, this vanishes by orthogonality.  If
they are the same, this is exactly the same expression we obtain for the
expectation value of the quark plus antiquark divergent mass operators; except
with the opposite sign. Therefore, there is a divergence in the quark-antiquark
interaction that is independent of their relative motion and which exactly cancels
the divergent masses!  These cancellations only occur for color singlets, and
they occur for any color singlet state with an arbitrary number of quarks and
gluons.  Moreover, these cancellations appear directly in the hamiltonian matrix
elements, so we can take the $\epsilon \rightarrow 0$ limit before diagonalizing
the matrix.

This is half of the simple confinement mechanism.  At this point it is
possible to obtain finite mass hadrons even though the parton masses
diverge.  However, since the cancellations are independent of the
relative parton motion, we must study the residual interactions to see
if they are confining.  Since I am interested in the long-range
interaction, I will study the fourier transform of the potential and
compute $V(r)-V(0)$ so that the divergent constant in which we are no
longer interested is canceled.

\begin{equation}
V_{singular}(r)-V_{singular}(0) \rightarrow 
{g_\Lambda^2 C_F \Lambda^2 \over 4
\pi^2 {\cal P}^+} \; \log\bigl(|x^-|\bigr) \;,
\end{equation}

\noindent when $x_\perp=0$ and $|x^-| \rightarrow \infty$; and

\begin{equation}
V_{singular}(r)-V_{singular}(0) \rightarrow 
{g_\Lambda^2 C_F \Lambda^2 \over 2
\pi^2 {\cal P}^+} \; \log\bigl(|x_\perp|\bigr) \;,
\end{equation}

\noindent when $|x_\perp| \rightarrow \infty$ and $x^-=0$.  This potential is not
rotationally symmetric,  but it diverges logarithmically in all directions.

If the potential is not rotationally symmetric, how can rotational symmetry be
restored?  In light-front field theory rotations are dynamical. While it may be
possible for rotational symmetry to be realized approximately in low-lying
quark-antiquark states, exact symmetry requires additional explicit partons and
even approximate rotational symmetry will require additional partons if we
study highly excited states.  We expect excited physical states in which a
quark and antiquark are separated by a large distance to contain gluons.  There
is no reason to assume that the gluon content of these states is the same when
the state is rotated, so rotational symmetry will be restored in highly excited
states only if we allow additional partons.  This complicates our attempt to
derive a constituent picture, but we only need the constituent picture to work
well for low-lying states.  The intermediate range part of the potential is
rotationally symmetric, and we may expect the ground state hadrons to be
dominated by the valence configuration.

Isn't the confining potential supposed to be linear and not logarithmic?  There
is no conclusive evidence that the long-range potential is linear, and heavy
quark phenomenology shows that a logarithmic potential can work quite well;
however, lattice calculations provide strong evidence for a linear potential. 
However, low-lying states are not sensitive to the longest-range part of the
interaction; and light quark-antiquark pairs prevent even exited states from
being sensitive to the longest-range part of the interaction. In any case, I do
not want to argue that these calculations show that the long-range potential in
light-front QCD is logarithmic.  Higher order corrections could produce powers
of logarithms that add up to produce a linear potential.

The above argument seems to apply directly to QED at first sight.  Is QED
confining?  There is a confining interaction between charged particles in the
hamiltonian, but there is no strong interaction between charged particles and
photons.  To see if confinement survives in QED we should include the confining
interaction in $H_0$, and then compute corrections in bound state perturbation
theory.  In QED the second order correction from photon exchange below the
cutoff exactly cancels confinement.  This implies that if confinement is
included in $H_0$, higher order corrections are large.  If the Coulomb
interaction, which appears in $H$ also, is included in $H_0$, bound state
perturbation theory appears to converge rapidly.  On the other hand, in QCD
gluons also experience a confining interaction.  When second order bound state
perturbation theory is used to study the effect of the exchange of confined
gluons, it is seen that gluon exchange does not cancel the confining interaction
in $H_0$; so this picture of confinement is at least self-consistent.

The important point is that $H$ contains a confining interaction that we are
free to include in $H_0$, giving us some hope of finding a reasonable bound
state perturbation theory for hadrons that resembles the bound state
perturbation theory that has been successfully applied to the study of atoms.

\vskip.15in
\noindent {\bf Summary}
\vskip.1in

A constituent picture of hadrons may emerge in QCD if we use:

\begin{itemize}

\item hamiltonian light-front field theory

\item a cutoff of order $\Lambda_{QCD}$ on energy changes that violates manifest
covariance and gauge invariance

\item a similarity renormalization group and coupling coherence

\end{itemize}

Bound states can be studied using bound state perturbation theory.  The
effective hamiltonian is computed as an expansion in the strong coupling, and
then divided into $H=H_0+V$, with $V$ treated perturbatively.  $H_0$ must
include all essential interactions.  We have found that $H$ contains an order
$\alpha$ logarithmically confining two-body interaction between all colored
partons, and we have begun studies of bound states using this confining
interaction, as discussed in the talk by Martina Brisudov{\'a}.

\vskip.15in
\noindent {\bf Acknowledgment}
\vskip.1in

I wish to thank the organizers for the opportunity to visit the beautiful
Slovak Republic. This talk could be viewed as an introduction to Martina
Brisudov{\'a}'s talk, which covered results for heavy mesons and I am indebted
to Martina for many insights.  I have profited from conversations with many
people, particularly Ken Wilson, Stan G{\l}azek, Brent Allen and Billy Jones. 
Finally I want to apologize to the many people whose work should have been
referenced.  This work was supported by the National Science Foundation under
grant PHY-9409042.

\vskip.15in
\noindent {\bf References}
\vskip.1in

\noindent
1. \refer{K. G. Wilson, T. S. Walhout, A. Harindranath, W.-M. Zhang, R. J.
Perry, St. D. G{\l}azek}{Phys. Rev. D}{49}{1994}{6720}

\noindent
2. R. J. Perry, in {\sl Proceedings of Hadrons 94}, World Scientific,
1995 (hep-th/9407056);

\noindent
3. K. G. Wilson and D. G. Robertson, in {\sl Proceedings of the Fourth
International Workshop on Light-Front Quantization and Non-Perturbative
Dynamics}, World Scientific, 1995 (hep-th/9411007);

\noindent
4. R. J. Perry, in {\sl Proceedings of the Fourth International Workshop on
Light-Front Quantization and Non-Perturbative Dynamics}, World Scientific, 1995
(hep-th/9411037);

\noindent
5. M. Brisudov{\'a} and R. Perry, {\sl Initial studies of bound states in
light-front QCD}, to appear in Phys. Rev. D, 1996 (hep-ph/9511443);

\noindent
6. \refer {R. J. Perry, A. Harindranath, and K. G. Wilson}{Phys. Rev. Lett.}{65}
{1990}{2959}

\noindent
7. \refer {R. J. Perry}{Ann. Phys.}{232}{1994}{116}

\noindent
8. \refer {St. D. G{\l}azek and K.G. Wilson}{Phys. Rev. D}{48}{1993}{5863}

\noindent
9. \refer {St. D. G{\l}azek and K.G. Wilson}{Phys. Rev. D}{49}{1994}{4214}

\noindent
10. \refer{R. J. Perry and K. G. Wilson}{Nucl. Phys. B} {403}{1993}{587}

\noindent
11. \refer{W. M. Zhang and A. Harindranath}{Phys. Rev. D}{48}{1993}{4868, 4881,
4903}

\end{document}